\documentclass[pra,twocolumn,showpacs,amsfonts,amsmath,floatfix]{revtex4}

\usepackage[pdftex]{graphicx}
\usepackage{array}
\usepackage{ulem}
\usepackage{url}

\begin{document}
\title{Analysis of the design of a passive resonant miniature optical gyroscope based on integrated optics technologies}

\author{Gilles Feugnet$^{1}$}
\author{Alexia Ravaille$^{1,2,3}$ }
\author{Sylvain Schwartz$^{1}$}
\author{Fabien Bretenaker$^{2}$}
\email{Fabien.Bretenaker@u-psud.fr}
 \affiliation {$^1$Thales Research \& Technology, 91120 Palaiseau, France}
 \affiliation {$^2$Laboratoire Aim\'e Cotton, Universit\'e Paris-Sud, ENS Paris-Saclay, CNRS, Universit\'e Paris-Saclay, Orsay, France}
 \affiliation {$^3$Thales Avionics, 40 Rue de la Brelandi\`ere, 86100 Ch\^atellerault, France}


\begin{abstract}
We present a simple analysis of the design of a passive miniature resonant optical gyroscope. By combining the requirements on the angular random walk and the bias stability, we end up with simple expressions of the minimum diameter of the ring waveguide cavity and the maximum power that should be used to probe it. Using state-of-the-art performances of photonic integrated circuit and whispering gallery mode technologies in terms of propagation losses and mode size, we show that tactical grade gyroscope performances can be achieved with a diameter of a few cm provided the detrimental influence of Kerr effect is mitigated, using for instance an active control of the unbalance in the intensities. We further extend the analysis to medium performance gyroscope and give some hints on the efforts to be made to potentially demonstrate a miniature resonant optical gyroscope with this level of performance.    \end{abstract}
\maketitle

\section{Introduction}
\label{intro} 
The fabrication of a navigation-grade miniature optical gyroscope has been the aim of an old quest. Indeed, the possible realization of a miniature optical gyroscope integrated on an optical chip and having a bias stability better than $1^{\circ}/\mathrm{h}$ could have a strong impact on the medium/high performance gyroscope market, currently dominated by well-established ring laser gyroscope\cite{Aronowitz1971} or interferometric fiber optical gyroscope\cite{Lefevre1993}. 
The Resonant Miniature Optical Gyroscope (RMOG), based on a waveguide-type ring resonator, is an attracting approach where the reduction of the optical path length is compensated by the cavity Q factor, as for other passive resonant devices\cite{Ezekiel1977}. Such RMOG becomes more and more realistic thanks to the progress of the Photonic Integrated Circuit (PIC) and Whispering Gallery Mode Resonator (WGMR) technologies. PIC technology, on the one hand, offers the potential of cost reduction through a collective manufacturing process like semiconductor industry and the generalization of the multi-wafer-run capability implemented by several foundries. There are now several platforms assisting end-users to design and manufacture PICs\cite{Jeppix,VLCPhotonics}. PIC development achieved a remarkable improvement of the losses\cite{Spencer2014,Yang2017}, heterogeneous integration of active indium phosphide gain sections or photodiodes with passive silicon or silicon nitride circuits\cite{Roelkens2013, Bowers2015}, and packaging\cite{Carroll2016} . On the other hand, WGMRs are not always fully compatible with collective manufacturing as they need to be diamond polished or they need to be heterogeneously reported on a substrate or coupled to a tapered fiber to inject and extract light. However, among the best performances of miniature cavities and gyroscope were demonstrated \cite{Liang2017} with this approach. 

In such a gyro architecture, the passive optical cavity is probed by two coherent counterpropagating waves, and the rotation rate is retrieved by monitoring the modifications of the resonant frequencies of the clockwise (CW) and counterclockwise (CCW) beams circulating inside the cavity. From the literature\cite{Ezekiel1977}, it is known that the partially reduced sensitivity of the RMOG could also be compensated by a higher laser incident power, in order to reduce shot noise. However, it is also known that a difference in the intensities of the CW and CCW beams induces via Kerr effect a non-reciprocal index difference \cite{Lefevre1993} resulting in a bias in the gyroscope response\cite{Iwatsuki1986}. Increasing the incident laser power then leads  to the necessity of a tighter control of this difference, which becomes increasingly challenging. In this paper, we conduct an analysis, that shows that it is possible, under realistic assumptions, to find an expression of the minimum cavity diameter and the maximum incident power the cavity should be probed with to meet a given requirement on both the angular random walk and bias stability of such a gyroscope. Our analysis takes into account i) the PIC and WGMR technology performances in term of propagation losses and mode size and ii) the difference of the counter propagating beam intensities. We then use this simple model to assess the feasibility of a RMOG for navigation applications.

The paper is organized as follows. In Section \ref{model}, we introduce our cavity model. In Section \ref{limits}, we use this model to derive the necessary minimum cavity diameter and the maximum incident power allowed to meet a set of navigation grade performances. As these values depend on the ring waveguide propagation losses, the Kerr effect coefficient, and the mode size, we conduct in Section \ref{applications} a parametric analysis using numbers from state-of-the-art PIC  and WGMR technology in order to assess  the feasibility of low (tactical) and medium performance RMOGs. We give some perspectives in terms of gyro development in Section \ref{conclusion}. These sections are supported by five appendices where we give the details of some calculations.  

\section{Cavity Model}
\label{model} 

\begin{figure}
\begin{center}
\begin{tabular}{c}
\includegraphics[height=5.5cm]{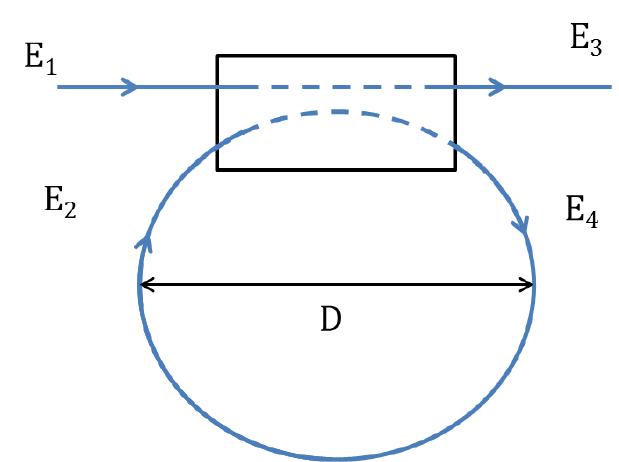}
\end{tabular}
\end{center} 
\caption 
{ \label{Fig01}
Schematics of the gyroscope sensing ring cavity. } 
\end{figure}

The rotation sensing ring cavity of diameter $D$ is schematized in Fig.\,\ref{Fig01}. We suppose that the cavity has only one coupler \cite{Ciminelli2013}. 

In order to maintain the laser at resonance, we implement a Pound Drever Hall (PDH) locking scheme \cite{Drever1983,Black2001}. We use the notations of Stokes et al.\cite{Stokes1982} by introducing the intensity losses $\gamma_0$ and the intensity coupling coefficient $\kappa$ of the coupler, which relate the coupler output and input intensities through the following relations:
\begin{align}
E_3&=\sqrt{1-\gamma_0}\left(\sqrt{1-\kappa}\,E_1+ i\sqrt{\kappa}\,E_2\right)\;,\label{eq02}\\
E_4&=\sqrt{1-\gamma_0}\left( i\sqrt{\kappa}\,E_1+ \sqrt{1-\kappa}\,E_2\right)\;,\label{eq03}
\end{align}
where the field complex amplitudes are labeled according to Fig.\,\ref{Fig01}. 

In appendix \ref{appendA}, we use these definitions to derive the expressions of the intracavity field and of the field reflected by the cavity. This leads to the following full width at half maximum for the cavity resonance dip:
\begin{equation}
\Delta\nu_{\mathrm{FWHM}}=\frac{1-\mathrm{e}^{-\alpha L/2}\sqrt{1-\gamma_0}\sqrt{1-\kappa}}{\pi}\Delta\nu_{\mathrm{FSR}}\equiv \frac{\Delta\nu_{\mathrm{FSR}}}{\mathcal{F}}\ ,\label{eq04}
\end{equation}
where $\Delta\nu_{\mathrm{FSR}}$ is the cavity free spectral range, $\alpha$ and $L$ are the waveguide intensity linear attenuation coefficient and length, and where we have defined the finesse $\mathcal{F}$ of the cavity. As discussed in appendix \ref{appendA} and in Ref.\;\citenum{DellOlio2015}, this expression is valid only in the low-loss approximation.

The noise properties of the PDH locking scheme can be optimized by maximizing the slope of the error signal at resonance. We show in Appendix \ref{appendB} that this optimum is obtained when the cavity obeys the so-called critical coupling condition \cite{Dumeige2008}, for which the resonance dip goes down to zero reflection, and which is obtained by equating the coupling factor with the internal losses of the resonator:
\begin{equation}
\sqrt{1-\kappa}=\sqrt{1-\gamma_0}\;\mathrm{e}^{-\alpha L/2}\ .\label{eq05}
\end{equation}
As shown in Appendix \ref{appendB}, optimizing the slope of the PDH error signal leads to a result different from the optimization of the cavity finesse, because the slope depends on both the finesse and the contrast of the resonances. In the case where the condition of Eq.\;(\ref{eq05}) is satisfied, Eq.\;(\ref{eq04}) becomes:
\begin{equation}
\frac{\Delta\nu_{\mathrm{FWHM}}}{\Delta\nu_{\mathrm{FSR}}}=\frac{1-\mathrm{e}^{-\alpha L}(1-\gamma_0)}{\pi}\ .\label{eq06}
\end{equation}

In order to enhance the sensitivity of our gyroscope, it is important that the losses be small. In this case, we can use the following approximation
\begin{equation}
\mathrm{e}^{-\alpha L}\simeq 1-\alpha L\ ,\label{eq07}
\end{equation}
and also suppose that the coupler losses $\gamma_0$ are much smaller than the propagation losses, leading to:
\begin{equation}
\kappa=1-(1-\gamma_0)\mathrm{e}^{-\alpha L}\simeq\alpha L\ .\label{eq08}
\end{equation}
Using Eqs.\,(\ref{eq05}), (\ref{eq06}), and (\ref{eq08}), this leads to:
\begin{equation}
\frac{\Delta\nu_{\mathrm{FWHM}}}{\Delta\nu_{\mathrm{FSR}}}=\frac{1}{\mathcal{F}}\simeq\frac{\alpha L}{\pi}=\alpha D\ ,\label{eq09}
\end{equation}
where $D$ is the diameter of the resonator (see Fig. \ref{Fig01}).

For a cavity under critical coupling conditions, we show in Appendix \ref{appendA} that the intracavity intensity $|E_4|^2$ at resonance is given by:
\begin{equation}
\left(\frac{|E_4|^2}{|E_1|^2}\right)_{\mathrm{resonance}}=\frac{1-\gamma_0}{1-(1-\gamma_0)\mathrm{e}^{-\alpha L}}\ .\label{eq10}
\end{equation} 
In the low loss approximation, we can use Eq. (\ref{eq08}) to obtain:
\begin{equation}
\left(\frac{|E_4|^2}{|E_1|^2}\right)_{\mathrm{resonance}}\simeq\frac{1}{\kappa}\simeq\frac{1}{\alpha\pi D}\ .\label{eq11}
\end{equation} 
This expression, already obtained in Ref. \citenum{Yariv2000}, stresses the direct link between the intracavity intensity and the propagation losses for a high finesse cavity in critical coupling regime.

\section{Shot Noise and Kerr Effect Related Performance Limits}
\label{limits}
We now focus on the ultimate achievable performances of the RMOG, assuming in the following sections that the only limitations are due to the shot noise and the Kerr effect. All other sources of noises such as the laser source or electronic noises, or limitations for instance from Rayleigh backscattering,\cite{Mao2011} or polarization noise\cite{Sanders1993,Ma2012}, are supposed to be mitigated. Of course, in a real development, these should be conveniently addressed, which is not a task that should be underestimated. We will come back to Brillouin scattering at the end of Sec. \ref{applications} to show that it is negligible with the materials that we consider. 
\subsection{Shot Noise Limit}
The smallest measurable angular velocity in a time $\tau$ depends on the slope of the PDH servo-locking slope, the Sagnac effect scale factor \cite{Sagnac1913}, and the shot noise level. Its expression is derived in Appendix \ref{appendC}, and reads:
\begin{equation}
\delta\dot{\theta}_{\mathrm{SNL}}=\frac{n_0\Delta\nu_{\mathrm{FWHM}}}{D}\sqrt{\frac{\lambda h c}{2\chi \tau P_0}}\ ,\label{eq12}
\end{equation}
where $n_0$ is the guided mode effective index, $\lambda$ the wavelength, $\chi$ the detector quantum efficiency, which we take equal to 1 in the following, $\tau$ the integration time, and $P_0$ the optical power incident on the cavity. The quantity $\delta\dot{\theta}_{\mathrm{SNL}}\sqrt{\tau}$ sets a limit on the minimal Angular Random Walk (ARW) of the device.

\subsection{Kerr Effect Induced Bias}
We then focus on the bias instability resulting from the Kerr effect. We show in Appendix \ref{appendD} that the Kerr effect introduces a rotation rate bias given by:
\begin{equation}
 \dot{\theta}_{\mathrm{Kerr}}=\frac{c\,n_2\,\mathcal{F}}{2\pi\sigma D}\Delta P_0\ ,\label{eq13}
\end{equation}
where $\Delta P_0$ is the difference between the clockwise and counterclockwise powers incident on the sensing cavity, $n_2$ the nonlinear refractive index of the waveguide, and $\sigma$ the guided mode area. It is important to notice that Eq.\;(\ref{eq13}) takes into account only the bias induced by the unbalance of the incident powers and not the extra noise induced by the differences between the frequencies or the voltages used for the PDH locking modulations in the two propagation directions.\cite{Iwatsuki1986,Sanders1993}

The bias of Eq.\;(\ref{eq13}) should be maintained below the maximum bias $ \dot{\theta}_{\mathrm{bias}}$ required by the gyro specifications, thus requiring a certain level of control of $\Delta P_0$. For example, for a silica cavity ($n_2=2.7\times10^{-20}\,\mathrm{m}^2/\mathrm{W}$\cite{Weber1978}) of finesse $\mathcal{F}=40$, diameter $D=25\,\mathrm{mm}$, and mode section area $\sigma=33\,\mu\mathrm{m}^2$, Eq.\,(\ref{eq13}) leads to a bias $\dot{\theta}_{\mathrm{Kerr}}/\Delta P_0=4\,^{\circ}/\mathrm{s}/\mathrm{mW}$, in fair agreement with the measured value of $5.3\,^{\circ}/\mathrm{s}/\mathrm{mW}$ reported in Ref.\,\citenum{Li2014}, after having taken into account the factor of two coming from the fact that only one half of the incident power is at the carrier frequency that is resonant with the cavity.

\subsection{Derivation of the Cavity Design Guidelines}
Equations (\ref{eq12}) and (\ref{eq13}) predict the ultimate performances of a RMOG depending on design parameters. Let us shortly summarize the hypothesis along which these expressions were obtained: i) The cavity exhibits low losses. The numerical results given latter will prove that this must be the case to reach the desired performances. ii) The coupler losses $\gamma_0$ are negligible compared to the propagation losses $\alpha L$. This is a reasonable assumption for state-of-the-art PICs couplers and waveguides \cite{Spencer2014}. However, this hypothesis may be slightly too optimistic for WGMRs. iii) A PDH locking-scheme is implemented and the phase modulation amplitude is set at its optimum\cite{Black2001} ($\beta=1.08\,\mathrm{rad}$). iv) The cavity is set at the critical coupling to optimize the PDH slope (see Appendix \ref{appendB}). Although this is not a critical parameter for our analysis, this can be a tricky issue in term of manufacturing as the propagation losses might be not totally reproducible or predictable. However, it was demonstrated in Refs.\,\citenum{Wang2013,Strain2015} that thermally controlled variable couplers can be integrated on a PIC, allowing post-fabrication tuning of the coupling efficiency, at the expense of potential drifts or errors if an active control is necessary. v) The detector efficiency $\chi$ is equal to 1 and the shot noise limit is reached.

Equations (\ref{eq12}) and (\ref{eq13}) contain different kinds of parameters. Some of them, such as $\lambda$, $c$, $h$, are physical constants that have fixed values. Some of them depend on the degree of maturity of the chosen technology. This is the case of $\alpha$, $n_2$, $\sigma$, $n_0$, that depend on the chosen PIC or WGMR technology, and of the degree to which one is able to balance the two couterpropagating intracavity powers, which we will parametrize by introducing the following notation
\begin{equation}
\xi=\frac{\Delta P}{P}=\frac{\Delta P_0}{P_0}\ ,\label{eq14}
\end{equation}
where $\Delta P/P$ and $\Delta P_0/P_0$ are the relative unbalance of the intracavity and incident powers, respectively. The smallest achievable value of the parameter $\xi$ depends on the efforts put into the active control of $\Delta P_0$ through relevant electronics controls. Finally, the two remaining parameters in Eqs.\;(\ref{eq12}) and (\ref{eq13}) are $D$ and $P_0$, which are the true design parameters of the gyro.

Consequently, our aim here is to derive the values of the design parameters $D$ and $P_0$ to achieve a given performance $ARW$ and $\dot{\theta}_{\mathrm{bias}}$, taking into account the parameters $\alpha$, $n_2$, $\sigma$, $n_0$ of the chosen technology. By combining Eqs. (\ref{eq12}) and (\ref{eq13}) with Eqs. (\ref{eq09}) and Eqs. (\ref{eq11}), we obtain the following constraints on $P_0$ and $D$:
\begin{align}
\frac{P_0}{D^2}<\frac{2\pi}{c}\ \frac{\sigma\alpha}{n_2}\ \frac{1}{\xi}\ \dot{\theta}_{\mathrm{bias}}\ ,\label{eq15}\\
\sqrt{P_0}D>\sqrt{\frac{\lambda h c^3}{2\pi^2}}\ \alpha\ \frac{1}{ARW}\ .\label{eq16}
\end{align}
One can see that these relations impose contradictory constraints on the input optical power $P_0$. In particular, for a fixed value of $D$, Eq. (\ref{eq15}) imposes a \textit{maximum} value of $P_0$ to limit the bias instabilities, which scales like $\alpha$, while Eq. (\ref{eq16}) imposes a \textit{minimum} value for $P_0$ to reach the requires ARW performance, which scales like $\alpha^2$.

\section{Applications to Tactical and Medium Performance Gyroscopes}
\label{applications} 
A tactical grade (i.e., low precision navigation grade) gyroscope usually has the following performance requirements:
\begin{align}
ARW& \leq 0.1\;^{\circ}/\sqrt{\mathrm{h}}=2.9\times10^{-5}\;\mathrm{rad}/\sqrt{\mathrm{s}}\ ,\label{eq17}\\
\dot{\theta}_{\mathrm{bias}}& \leq 1\;^{\circ}/\mathrm{h}=4.8\times10^{-6}\;\mathrm{rad}/\mathrm{s}\ ,\label{eq18}
\end{align}
while for a medium precision gyroscope these specifications become:
\begin{align}
ARW& \leq 0.01\;^{\circ}/\sqrt{\mathrm{h}}=2.9\times10^{-6}\;\mathrm{rad}/\sqrt{\mathrm{s}}\ ,\label{eq19}\\
\dot{\theta}_{\mathrm{bias}}& \leq 0.1\;^{\circ}/\mathrm{h}=4.8\times10^{-7}\;\mathrm{rad}/\mathrm{s}\ .\label{eq20}
\end{align}
Given those specifications, the question now is to decide which PIC or WGMR technology is the most favorable to build a RMOG that could meet them. Table \ref{table01} summarizes the propagation losses and nonlinear index of four different PIC technologies and one WGMR technology, namely silicon on insulator (SOI, see Refs.\;\citenum{Jalali2006,Thomson2016}), indium phosphide (InP, see Refs.\;\citenum{Shih2012,Gilardi2014,DAgostino2015,Ciminelli2016}), silicon nitride (SiN, see Refs.\;\citenum{Worhoff2015,Kruckel2015}), silicon-chip-based  monolithic silica resonators (SiO$_2$, see Refs.\;\citenum{Lee2012,Lee2017,Yang2017}), and CaF$_2$ WGMRs which are heterogeneously reported on a micro-optical chip \cite{Liang2017}. 

Note also that we suppose that the performances reported in Table \ref{table01} were all obtained for single-mode waveguides, which may not always be the case.

\begin{table*}[ht]
\hspace{4.0cm}
\caption{Propagation losses $\alpha$ and non-linear index of refraction $n_2$ for five different PIC materials.} 
\label{table01}
\begin{center}       
\begin{tabular}{|c|c|c|} 
\hline
\rule[-1ex]{0pt}{3.5ex}  Technology & $\alpha$ & $n_2$  \\
\hline\hline
\rule[-1ex]{0pt}{3.5ex}  SOI & 2.7 dB/m (Ref.\;\citenum{Biberman2012}) & $5\times10^{-18}$ m$^2$/W (Ref.\;\citenum{Bristow2007}) \\
\hline
\rule[-1ex]{0pt}{3.5ex}  InP & 0.35 dB/cm (Refs.\;\citenum{DAgostino2015,Ciminelli2016}) & $10^{-16}$ m$^2$/W (Ref.\;\citenum{Leuthold1998}) \\
\hline
\rule[-1ex]{0pt}{3.5ex}  SiN & 0.32 dB/m (Ref.\;\citenum{Spencer2014}) & $2.4\times10^{-19}$ m$^2$/W (Ref.\;\citenum{Tan2010}) \\
\hline
\rule[-1ex]{0pt}{3.5ex}  SiO$_2$ & 0.11 dB/m (Ref.\;\citenum{Yang2017}) & $2.7\times10^{-20}$ m$^2$/W (Ref.\;\citenum{Weber1978}) \\
\hline
\rule[-1ex]{0pt}{3.5ex}  CaF$_2$ & 0.0016 dB/m (Ref.\;\citenum{Liang2017}) &$3.6\times10^{-20}$ m$^2$/W (Ref.\;\citenum{Savchenkov2004}) \\
\hline
\end{tabular}
\end{center}
\end{table*} 

From Table \ref{table01}, the most promising technologies seem to be the ones based on SiN, SiO$_2$, and CaF$_2$ for two reasons. First, these materials are the one with which the lowest propagation losses were demonstrated. Second, their nonlinear indices of refraction $n_2$ are smaller than the other materials mentioned in Table \ref{table01}.  Actually, SiN is even more favorable than what can be seen in this table because its \textit{effective nonlinear index of refraction} can be made quite close to the one of silica. Indeed, because the contrast between the refractive indices of SiN and SiO$_2$, which is used as the substrate for the PIC, is pretty small, the mode profile can be tailored to be only weakly confined so that the field mainly propagates inside SiO$_2$. Actually, as can be seen in Ref.\;\citenum{Spencer2014}, the confinement factor $\eta$, defined as the fraction of the mode power that propagates in the central SiN core \cite{Yariv1989}, is only 0.03, meaning that the effective non-linear index is approximately given by
\begin{equation}
n_{2,\mathrm{eff}}=\eta\;n_2(\mathrm{SiN})+(1-\eta)\;n_2(\mathrm{SiO}_2)\approx3.4\times10^{-20}\;\mathrm{m}^2/\mathrm{W}\ .\label{eq21}
\end{equation}
Moreover, with such a low confinement, the mode area $\sigma$ is equal to $33\;\mu\mathrm{m}^2$ (see Ref.\;\citenum{Spencer2014}), thus further decreasing the bias induced by the Kerr effect. The mode diameter in SiO$_2$ resonators is taken to be equal to $37\;\mu\mathrm{m}^2$ (see Ref.\;\citenum{Lee2017}). The CaF$_2$ WGMR technology benefits from an even larger mode diameter, namely $\sigma=190\;\mu\mathrm{m}^2$, which we deduced from Ref. \citenum{Savchenkov2004}.

From Table \ref{table01}, SOI could also have been a possible candidate, provided some improvement on the propagation losses would be achievable. However, because the index of refraction contrast between the silicon layer and the SiO$_2$ substrate is very high, the mode remains mainly confined inside the silicon layer, meaning that the effective Kerr effect is close to the one of the bulk material reported in Table \ref{table01}, i. e., 100 times larger than that of SiO$_2$ and 10 times that of SiN (for further comparison between SOI and SiN, see for instance Ref. \citenum{Baets2016}). Concerning InP, it is also worth mentioning that other non-linear effects such as the two-photon absorption could become detrimental to the gyro performance even before the Kerr effect itself becomes a problem. 

\subsection{Tactical Grade Gyroscope}
\label{tactical}

\begin{figure*}
\begin{center}
\begin{tabular}{c}
\includegraphics[width=.8\textwidth]{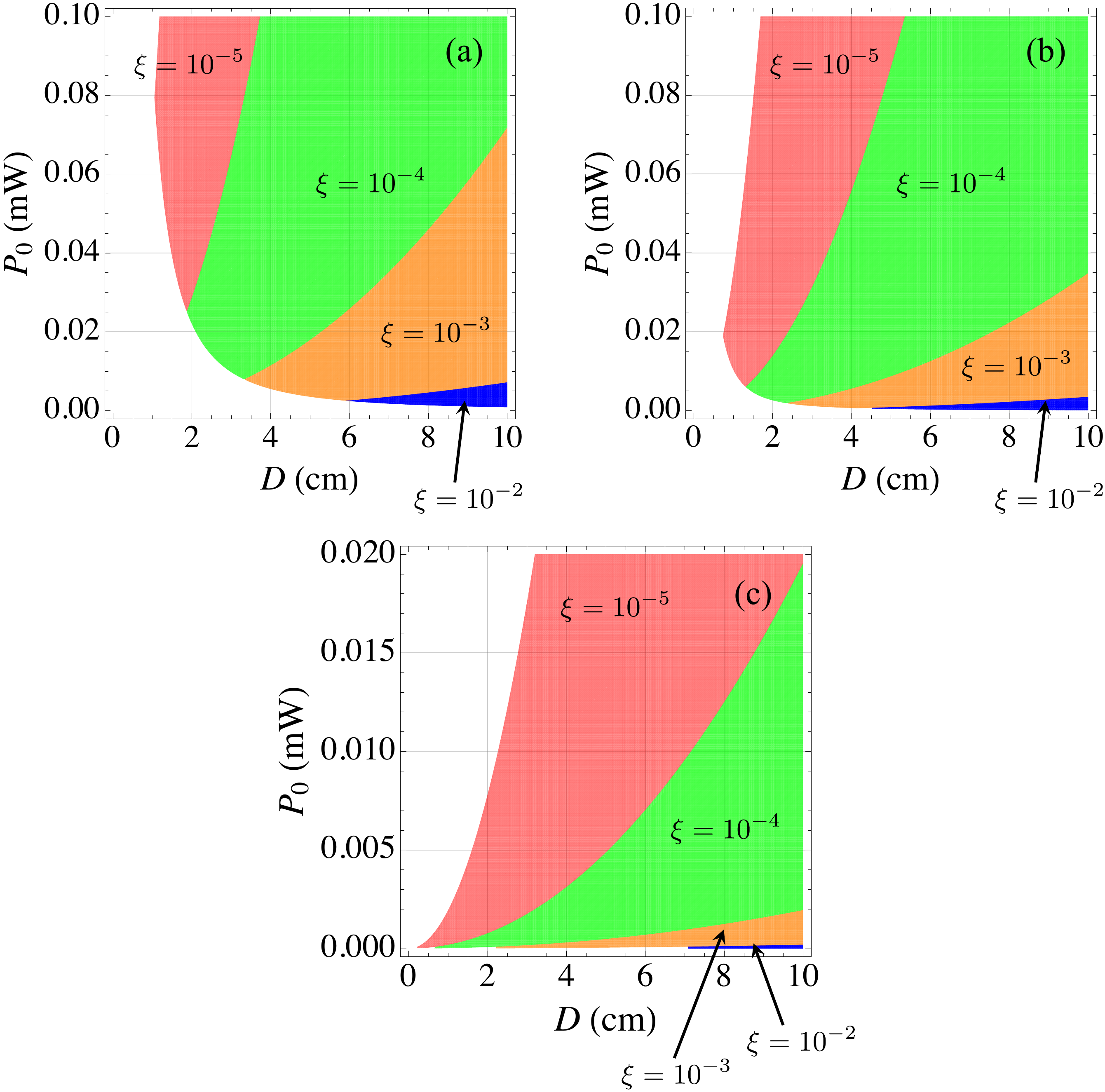}
\end{tabular}
\end{center} 
\caption 
{ \label{Fig02}
Areas in which Eqs. (\ref{eq15}) and (\ref{eq16}) are valid for a tactical grade gyroscope [see Eqs. (\ref{eq17}) and (\ref{eq18})] for(a)~SiN, (b) SiO$_2$, and (c) CaF$_2$ micro resonators. The values of the parameters are (a) $n_2=3.4\times10^{-20}\;\mathrm{m}^2/\mathrm{W}$, $\alpha=0.32\;\mathrm{dB/m}$, and $\sigma=33\;\mu\mathrm{m}^2$, (b) $n_2=2.7\times10^{-20}\;\mathrm{m}^2/\mathrm{W}$, $\alpha=0.11\;\mathrm{dB/m}$, and $\sigma=37\;\mu\mathrm{m}^2$ , and (c) $n_2=3.6\times10^{-20}\;\mathrm{m}^2/\mathrm{W}$, $\alpha=0.0016\;\mathrm{dB/m}$, and $\sigma=190\;\mu\mathrm{m}^2$. The four areas correspond to $\xi= 10^{-2}$, $10^{-3}$, $10^{-4}$, and $10^{-5}$. } 
\end{figure*}

With the figures we have obtained for SiN and SiO$_2$ PICs and CaF$_2$ WGMRs, respectively, Figs.\;\ref{Fig02}(a), \ref{Fig02}(b), and \ref{Fig02}(c) show the range of the parameters $D$ and $P_0$ for which the limit is compatible with tactical performances [see Eqs. (\ref{eq17}) and (\ref{eq18})] for four values of the maximum power imbalance $\xi$. These areas are obtained using the conditions given by Eqs. (\ref{eq15}) and (\ref{eq16}). 

The first thing we can notice by comparing Figs.\ \ref{Fig02}(a) and \ref{Fig02}(b) is that the results are quite similar for the two considered PIC materials. Looking closer into details, one can see that for $\xi=10^{-2}$, the diameter of the gyro must be larger than 6\;cm or 4.5\;cm in the cases of SiN and SiO$_2$, respectively. These are no longer really miniature dimensions. Besides, with such a value of $\xi$, the optical power must reduced to a few $\mu$W to mitigate the Kerr effect bias. Only a power imbalance control as good as $\xi=10^{-3}$, which is achievable with a servo-loop control, can permit to reduce $D$ below 4\;cm in the case of SiN and below 3\;cm in the case of SiO$_2$, with an optical power of the order of $10~\mu\mathrm{W}$. A power imbalance control as good as $\xi=10^{-4}$ could permit to decrease the cavity diameter down to 2 and 1.5\;cm for SiN and SiO$_2$, respectively.

From Fig.\ \ref{Fig02}(c), one could \textit{a priori} believe that CaF$_2$ WGMR can permit to achieve tactical grade performances with smaller dimensions than PIC technologies. However, this impression must be mitigated by several observations: i) the level of power needed to control the Kerr effect induced bias becomes so low, in the range of $1\;\mu$W, that detection noise problems may become an issue; ii) the calculations of Fig.\ \ref{Fig02}(c) have been performed by assuming critical coupling, which is far from being the case in real implementations like the one of Ref. \citenum{Liang2017}, where coupling losses are sixteen times larger than internal losses; iii) the tapered fiber coupling technique used for WGMR probably induces coupler losses, that are neglected here, which will further degrade the performance.

As a partial conclusion, Fig.\ \ref{Fig02} stresses the fact that one really needs to take into account the role of the Kerr effect, and not only the shot noise limit, in the achievable ultimate performance of the gyro. They also show that a control of the power imbalance between the two counter propagating curves is unavoidable.

\subsection{Medium Performance Gyroscope}
\label{tactical} 

\begin{figure*}[]
\begin{center}
\begin{tabular}{c}
\includegraphics[width=.8\textwidth]{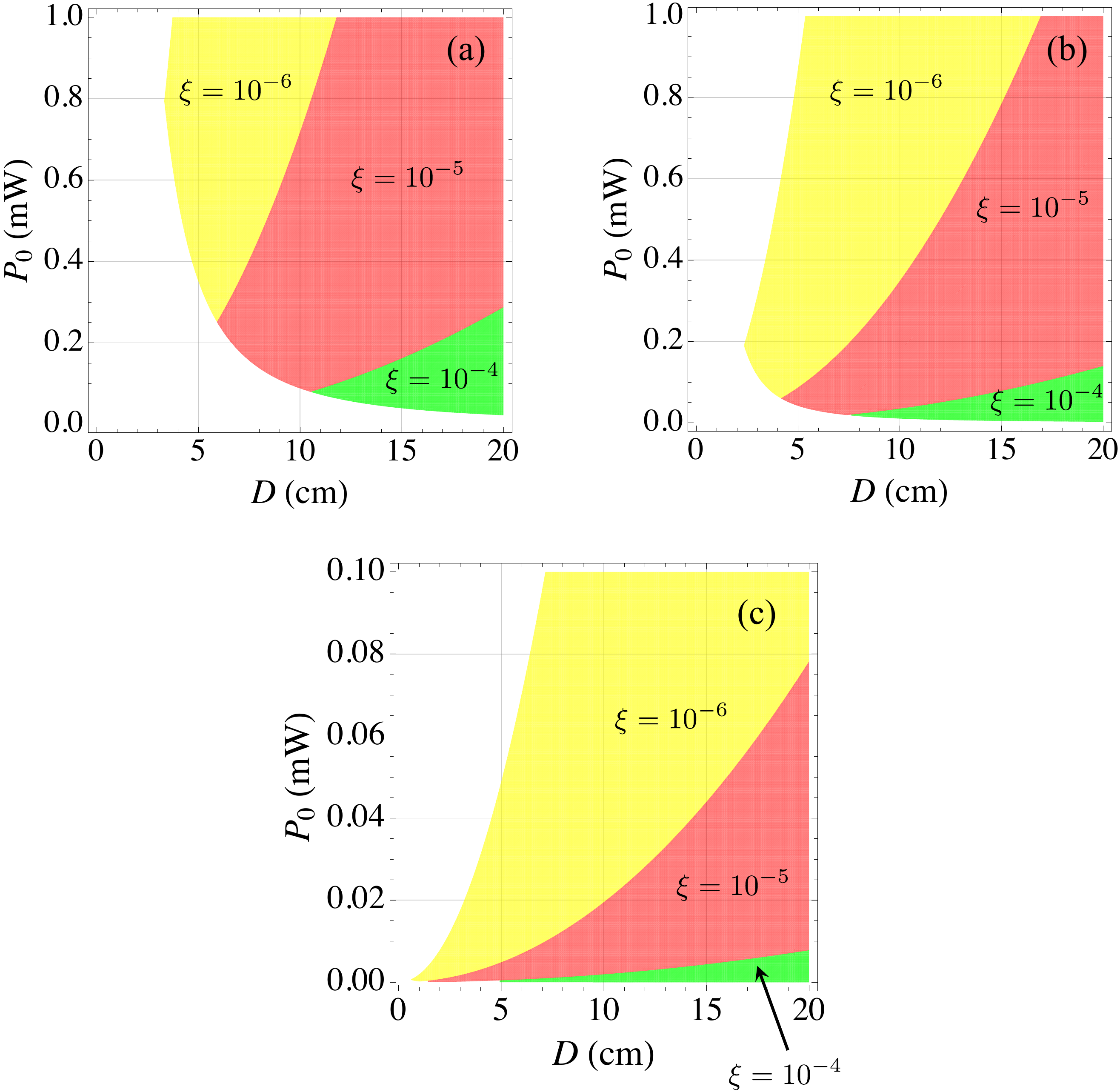}
\end{tabular}
\end{center} 
\caption 
{ \label{Fig03}
Areas in which Eqs. (\ref{eq15}) and (\ref{eq16}) are valid for a medium performance gyroscope [see Eqs. (\ref{eq19}) and (\ref{eq20})] for (a)~SiN, (b) SiO$_2$, and (c) CaF$_2$ micro resonators. The values of the parameters are (a) $n_2=3.4\times10^{-20}\;\mathrm{m}^2/\mathrm{W}$, $\alpha=0.32\;\mathrm{dB/m}$, and $\sigma=33\;\mu\mathrm{m}^2$, (b) $n_2=2.7\times10^{-20}\;\mathrm{m}^2/\mathrm{W}$, $\alpha=0.11\;\mathrm{dB/m}$, and $\sigma=37\;\mu\mathrm{m}^2$ , and (c) $n_2=3.6\times10^{-20}\;\mathrm{m}^2/\mathrm{W}$, $\alpha=0.0016\;\mathrm{dB/m}$, and $\sigma=190\;\mu\mathrm{m}^2$. The four areas correspond to $\xi= 10^{-4}$, $10^{-5}$, and $10^{-6}$.}
\end{figure*}

The situation is even worse in the case of a medium performance gyroscope, i. e. with the performance specifications given by Eqs. (\ref{eq19}) and (\ref{eq20}). As can be seen in Fig.\;\ref{Fig03}, an active control of the power imbalance as good as $\xi= 10^{-5}$ can only allow to reduce the minimum cavity diameter down to 6\;cm in the case of SiN and to 5\;cm in the case of SiO$_2$. From Fig.\;\ref{Fig03}(c), it seems possible to reduced this diameter down to 3 or 4 cm in the case of CaF$_2$ WGMRs. But the same discussion on the validity of the hypothesis as in the discussion on tactical performance gyroscopes applies here also.

With SiN and SiO$_2$ PIC technologies respectively, a tremendously precise power control ($\xi= 10^{-6}$) is necessary to obtain the desired performance for $D = 5\;\mathrm{cm}$ and $D = 3\;\mathrm{cm}$, as can be seen from Figs.\;\ref{Fig03}(a) and \ref{Fig03}(b). Fig.\;\ref{Fig03}(c) suggests that this diameter could only be slightly reduced by the use of a CaF$_2$ WGMR.

With the materials that we have chosen, and with the typical diameters and optical powers that we consider, one can check that Brillouin scattering is negligible, as shown in Appendix \ref{Brillouin}.

\subsection{Discussion}
The main conclusion of this simple analysis is that the Kerr effect deeply impacts the RMOG design as it rapidly limits the level of power allowed to probe the cavity. For such a given power limit, the only way to improve the sensitivity at the shot noise limit is to increase the cavity diameter, leading to cavity dimensions that become comparable to other optical gyro technologies. Even for the tactical grade gyroscope, which has the less stringent performances, relative power differences well below the percent level are required. With a power imbalance level in the $10^{-3}$ to $10^{-4}$ range, a miniature tactical grade gyroscope seems achievable with a ring cavity diameter in the few cm range. Making a medium performance RMOG seems much more problematic because the relative difference of power should be below $10^{-5}$,  not even mentioning the fact that the cavity should remain monomode with a single polarization or the manufacturing and cost issues associated with such a large diameter resonator.

In view of resonant gyro applications, PIC or WGMR design should consequently focus not only on the propagation losses but also on the mode area $\sigma$ in order to decrease the effective Kerr effect. Even with a reduction of the Kerr effect, operation of a truly miniature ring with a diameter of a few cm implies a mandatory active control of the intensities. Indeed, such a level of power imbalance control ($10^{-5}$ for a 3 cm diameter tactical grade diameter RMOG for instance) seems impossible to achieve in a passive way. In this respect, the demonstration in Ref. \citenum{Ma2014} of relative power imbalance actively reduced down to $2.5\times10^{-5}$ opens the way to such controls.  The intensity modulation scheme proposed in Ref. \citenum{Takiguchi1992} is also well adapted to the small cavities that we consider here, contrary to the one described in  Ref. \citenum{Iwatsuki1986}.

To obtain the above results, we have assumed that the coupler losses, $\gamma_0$, are negligible compared to the propagation losses $\alpha L$ and the coupler transmission $\kappa$. This assumption should be reassessed for the very small coupling values in the case of PIC resonators and may be far from being valid in the case of WGMRs, as discussed above. This would make things worse if this hypothesis is no longer true. Indeed, some extra coupler losses would lead to an increase of the cavity linewidth and smaller PDH slope as can be seen from Eqs.\;(\ref{eqB10}-\ref{eqB12}) in Appendix \ref{appendB}. The effect of these extra losses on the PDH slope should then be compensated for by increasing the incident power and/or the gyroscope resonator diameter.  We have also assumed that the cavity is exactly tuned to the critical coupling regime. This does not only make all the calculations simpler, allowing us to derive formulas that can be used as simple design rules of a RMOG (and actually for any resonant optical gyroscope) but it also optimizes the PDH slope. Thus, should the cavity coupling be set at a different regime, this would also lead to a need to increase the incident power or/and the gyroscope size to meet the desired performance limits.

Moreover, we assumed throughout this paper that the phase modulation depth was optimal for the PDH locking (i.e. 1.08 rad). However, it was shown\cite{Mao2011} that a modulation depth of 2.4 rad should be used in order to suppress the carrier and thus reduce the effect of the backscattering that we mentioned as a noise to be addressed. However, this modulation scheme leads to a somehow less steep error signal because the dominant terms are now proportional to $J_1(\beta)J_2(\beta)$, as shown in Ref.\;\citenum{Zhang2006}. A way to keep a modulation depth equal 1.08 rad while addressing the backscattering issue could be to interrogate the cavity with three different frequencies separated by an integer number of cavity free spectral ranges as proposed in Ref. \citenum{Schwartz2015}.

\section{Conclusion}
\label{conclusion} 
In conclusion, we have performed a simple analysis of a PIC-based or WGMR-based Resonant Miniature Optical Gyroscope with few realistic assumptions, namely critical coupling, high-Q cavity, negligible coupling losses, and Pound-Drever-Hall locking scheme driven with the optimal modulation depth. We have derived design rules to calculate the minimum gyroscope diameter and the maximum power to probe it, taking the shot noise limit and the Kerr-effect induced bias into account. From this, we conclude that the Kerr effect has a deep impact on the design of the gyroscope, and that the cavity Q-factor is not the unique parameter governing the gyro performance. In order to meet the bias stability requirement, Kerr effect puts a limit on the acceptable difference between the powers of the two counter-propagating probing beams. With a fixed given relative power imbalance, this drastically limits the maximum input power. It then becomes necessary to increase the gyroscope dimensions to fulfill the ARW requirement. 

More precisely, we conclude that even the goal of building a cm-diameter-scale tactical grade gyroscope (i.e. a relatively low performance grade in term of inertial navigation) already puts a strong constraint on the power balance, making it very challenging to build and raising cost related issues. Assuming some of the best demonstrated PIC or WGMR technologies so far, reaching the performances of such a gyroscope requires a mitigation of the Kerr effect, for instance with an active control of the counter propagating beam intensities. Second, for the same reasons,  important improvements on the technology, such as a strong decrease of the losses, would be necessary to make a medium grade RMOG feasible. A more radical approach could be to get rid of the Kerr effect itself by making the light propagate mainly in air as with hollow-core fibres \cite{Fsaifes2017}, by exploiting circular hollow core waveguides\cite{Kumar2010,Yang2012}, by adapting the slot-waveguide approach \cite{Almeida2004,Carlborg2010} to air core, or by pushing to its limits the wedge resonator geometry \cite{Lee2012,Yang2017} so that the sharp part of the waveguide is so thin that the mode confinement drastically decreases and the field mainly propagates in air. Another approach to reduce the Kerr effect limitation would be to identify new modulation solutions adapted to small cavities. 

Finally, let us mention that another approach\cite{Li2017}, based on an active Brillouin resonator, seems also promising to achieve a performant miniature optical gyroscope.

\appendix
\section{Cavity Resonance and Linewidth }
\label{appendA}
In this appendix, we derive the expression of the field reflected by and inside the cavity, in order to obtain Eqs. (\ref{eq04}-\ref{eq11}). Inside the cavity (see Fig. \ref{Fig01}), the fields $E_2$ and $E_4$ are related by:
\begin{equation}
E_2=E_4\mathrm{e}^{-\alpha L/2}\mathrm{e}^{-\mathrm{i}\Phi}\ ,\label{eqA01}
\end{equation}
where the round-trip phase shift is given by
\begin{equation}
\Phi=\frac{2\pi\nu n_0 L}{c}=\frac{\omega}{\Delta\nu_{\mathrm{FSR}}}\ ,\label{eqA02}
\end{equation}
where $\omega$ is the light angular frequency and $n_0$ the guided mode effective index. By combining this equation with Eq. (\ref{eq03}), we get
\begin{equation}
E_4=\frac{i\sqrt{1-\gamma_0}\sqrt{\kappa}}{1-\mathrm{e}^{-\alpha L/2}\mathrm{e}^{-\mathrm{i}\Phi}\sqrt{1-\gamma_0}\sqrt{1-\kappa}}E_1\ .\label{eqA03}
\end{equation}
At resonance, i. e., $\mathrm{e}^{-\mathrm{i}\Phi}=1$, the intracavity intensity is maximized and is given by:
\begin{equation}
\left(\frac{|E_4|^2}{|E_1|^2}\right)_{\mathrm{resonance}}=\frac{\kappa(1-\gamma_0)}{\left(1-\mathrm{e}^{-\alpha L/2}\sqrt{1-\gamma_0}\sqrt{1-\kappa}\right)^2}\ .\label{eqA04}
\end{equation} 
In the critical coupling regime given by Eq. (\ref{eq05}), we retrieve the expression of Eq.\ (\ref{eq10}).

The field reflected by the cavity is obtained by injecting Eqs. (\ref{eqA01}) and (\ref{eqA03}) into Eq. (\ref{eq02}), leading to:
\begin{equation}
E_3=\sqrt{1-\gamma_0}\left[\frac{\sqrt{1-\kappa}-\mathrm{e}^{-\alpha L/2}\mathrm{e}^{-i\Phi}\sqrt{1-\gamma_0}}{1-\mathrm{e}^{-\alpha L/2}\mathrm{e}^{-i\Phi}\sqrt{1-\gamma_0}\sqrt{1-\kappa}}\right]E_1\ .\label{eqA05}
\end{equation}
\begin{widetext}
This leads to the following expression for the reflected intensity:
\begin{equation}
\frac{|E_3|^2}{|E_1|^2}=(1-\gamma_0)\left[1-\kappa\frac{1-(1-\gamma_0)\mathrm{e}^{-\alpha L}}{\left(1-\mathrm{e}^{-\alpha L/2}\sqrt{1-\gamma_0}\sqrt{1-\kappa}\right)^2+4\mathrm{e}^{-\alpha L/2}\sqrt{1-\gamma_0}\sqrt{1-\kappa}\sin^2\Phi/2}\right]\ .\label{eqA06}
\end{equation}
\end{widetext}
The reflected intensity is minimum at resonance, i.e. for $\sin^2\Phi/2=0$, and maximum at anti-resonance, i.e. for $\sin^2\Phi/2=1$. The half-width at half-maximum of the resonance corresponds to the value 
\begin{equation}
\Phi_{1/2}=\pi\frac{\Delta\nu_{\mathrm{FWHM}}}{\Delta\nu_{\mathrm{FSR}}}\ \label{eqA07}
\end{equation}
 of the phase deviation with respect to resonance for which the reflected intensity is equal to the average of the minimum and maximum reflected intensities. From Eq. (\ref{eqA06}), it is given by:
\begin{equation}
\sin^2 \frac{\Phi_{1/2}}{2}=\frac{\left(1-\mathrm{e}^{-\alpha L}\sqrt{1-\gamma_0}\sqrt{1-\kappa}\right)^2}{2\left[1+\mathrm{e}^{-\alpha L}(1-\gamma_0)(1-\kappa)\right]}\ .\label{eqA08}
\end{equation}
In the case where the cavity finesse is large, one has $\sin \Phi_{1/2}\simeq \Phi_{1/2}$, leading to:
\begin{equation}
\frac{\Delta\nu_{\mathrm{FWHM}}}{\Delta\nu_{\mathrm{FSR}}}\simeq\frac{\sqrt{2}}{\pi}\frac{1-\mathrm{e}^{-\alpha L}\sqrt{1-\gamma_0}\sqrt{1-\kappa}}{\left[1+\mathrm{e}^{-\alpha L}(1-\gamma_0)(1-\kappa)\right]^{1/2}}\ \label{eqA09}
\end{equation}
For such a high finesse cavity, one has $1+\mathrm{e}^{-\alpha L}(1-\gamma_0)(1-\kappa)\simeq 2$, leading to Eq. (\ref{eq04}). In the case where the losses are too large for this approximation to be valid, an exact expression of the cavity linewidth can be found in Ref.\;\citenum{DellOlio2015}.

\section{Optimization of the Slope of the PDH Error Signal.}
\label{appendB}
In this appendix, we derive the slope of the PDH error signal for a single coupler ring cavity such as the one of Fig.\;\ref{Fig01} and show that the critical coupling is the one that maximizes this slope. We adopt the notations of Ref. \citenum{Black2001} where the incident field at angular frequency $\omega=2\pi\nu$, of complex amplitude $E_0$, is phase modulated at angular frequency $\Omega$ with an amplitude $\beta$. The incident field thus reads
\begin{equation}
E_0\mathrm{e}^{i\left[\omega t+\beta\sin(\Omega t)\right]}=E_0\sum_{n=-\infty}^{n=\infty}J_n(\beta)\mathrm{e}^{i\left(\omega+n\Omega\right)t}\ ,\label{eqB01}
\end{equation}
where $J_n$ is the Bessel function of order $n$. 
\begin{widetext}
The reflected field, limited to the central carrier and the first two sidebands is equal to :
\begin{equation}
E_0\left[F(\omega)J_0(\beta)\mathrm{e}^{i\omega t}+F(\omega+\Omega)J_1(\beta)\mathrm{e}^{i(\omega+\Omega)t}+F(\omega-\Omega)J_{-1}(\beta)\mathrm{e}^{i(\omega-\Omega) t}\right]\ ,\label{eqB02}
\end{equation}
where $F(\omega)$ is the cavity amplitude reflection coefficient given in Eq.\ (\ref{eqA05}):
\begin{equation}
F(\omega)=\sqrt{1-\gamma_0}\left[\frac{\sqrt{1-\kappa}-\mathrm{e}^{-\alpha L/2}\mathrm{e}^{-i\omega/\Delta\nu_{\mathrm{FSR}}}\sqrt{1-\gamma_0}}{1-\mathrm{e}^{-\alpha L/2}\mathrm{e}^{-i\omega/\Delta\nu_{\mathrm{FSR}}}\sqrt{1-\gamma_0}\sqrt{1-\kappa}}\right]\ .\label{eqB03}
\end{equation}
\end{widetext}
If we suppose that the sidebands are completely out of resonance, then $F(\omega\pm\Omega)\simeq-1$ and, following Ref. \citenum{Black2001}, the error signal after demodulation, i. e., multiplication by $\sin(\Omega t)$, and low-pass filtering is given by
\begin{equation}
\varepsilon(\omega)=2\,GP_0 \left|J_0(\beta)J_1(\beta)\right|\mathrm{Im}[F(\omega)]\ ,\label{eqB04}
\end{equation}
where $G$ is the optical to electrical conversion gain and $P_0$ the optical power associated with the incident laser field $E_0$. 
\begin{widetext}
The error signal is thus proportional to the imaginary part of $F(\omega)$, which, according to Eq.\;(\ref{eqB03}), is given by:
\begin{equation}
\mathrm{Im}[F(\omega)]=\frac{\kappa(1-\gamma_0)\mathrm{e}^{-\alpha L/2}\sin(\omega/\Delta\nu_{\mathrm{FSR}})}{\left(1-\mathrm{e}^{-\alpha L/2}\sqrt{1-\kappa}\sqrt{1-\gamma_0}\right)^2+4\mathrm{e}^{-\alpha L/2}\sqrt{1-\kappa}\sqrt{1-\gamma_0}\sin^2(\omega/2\Delta\nu_{\mathrm{FSR}})}\ .\label{eqB05}
\end{equation}
If we call $\delta\omega$ the shift of $\omega$ with respect to resonance and suppose that $\delta\omega\ll 2\pi\Delta\nu_{\mathrm{FSR}}$, then Eq.\;(\ref{eqB05}) becomes:
\begin{equation}
\mathrm{Im}[F(\delta\omega)]\simeq\frac{\kappa(1-\gamma_0)\mathrm{e}^{-\alpha L/2}}{\left(1-\mathrm{e}^{-\alpha L/2}\sqrt{1-\kappa}\sqrt{1-\gamma_0}\right)^2}\frac{\delta\omega}{\Delta\nu_{\mathrm{FSR}}}\ .\label{eqB06}
\end{equation}
\end{widetext}
Close to resonance, the evolution of the error signal with the cavity detuning is thus linear. One can optimize the slope of this evolution by chosing the value of the coupling $\kappa$ that satisfies
\begin{equation}
\frac{\mathrm{d}}{\mathrm{d}\kappa}\mathrm{Im}[F(\delta\omega)]=0\ .\label{eqB07}
\end{equation}
By taking the derivative of Eq.\;(\ref{eqB06}) with respect to $\kappa$, one retrieves exactly the condition (\ref{eq05}) for critical coupling:
\begin{equation}
\kappa=1-\mathrm{e}^{-\alpha L}(1-\gamma_0)\ .\label{eqB08}
\end{equation}
In the case of critical coupling, Eq.\,(\ref{eqB06}) becomes
\begin{equation}
\mathrm{Im}[F(\delta\omega)]\simeq\frac{(1-\gamma_0)\mathrm{e}^{-\alpha L/2}}{1-\mathrm{e}^{-\alpha L}(1-\gamma_0)}\frac{\delta\omega}{\Delta\nu_{\mathrm{FSR}}}\ .\label{eqB09}
\end{equation}
In the case of a high finesse cavity, we have $\mathrm{e}^{-\alpha L}(1-\gamma_0)\simeq 1-\alpha L-\gamma_0$, so that
\begin{equation}
\mathrm{Im}[F(\delta\omega)]\simeq\frac{1}{\alpha L+\gamma_0}\frac{\delta\omega}{\Delta\nu_{\mathrm{FSR}}}\ .\label{eqB09}
\end{equation}
Besides, Eq.\,(\ref{eq06}) leads to
\begin{equation}
\frac{1}{\mathcal{F}}=\frac{\Delta\nu_{\mathrm{FSR}}}{\Delta\nu_{\mathrm{FWHM}}}\simeq\frac{\alpha L+\gamma_0}{\pi}\ ,\label{eqB10}
\end{equation}
allowing us to rewrite Eq.\,(\ref{eqB09}) in the following form:
\begin{equation}
\mathrm{Im}[F(\delta\omega)]\simeq\frac{\delta\omega}{\pi\Delta\nu_{\mathrm{FWHM}}}\ .\label{eqB11}
\end{equation}
Using Eq.\;(\ref{eqB04}), we obtain finally:
\begin{equation}
\varepsilon(\delta\omega)=2\,GP_0 \left|J_0(\beta)J_1(\beta)\right|\frac{\delta\omega}{\pi\Delta\nu_{\mathrm{FWHM}}}\ .\label{eqB12}
\end{equation}

\section{Shot Noise Limit }
\label{appendC}
In this appendix, we use the slope error signal expression obtained in Appendix B to derive the shot noise limit of a resonant gyroscope. The detection of the power $P_{\mathrm{det}}$ reflected by the cavity creates a photocurrent given by 
\begin{equation}
i_{\mathrm{det}}=\frac{\chi e \lambda}{h c}P_{\mathrm{det}}\ ,\label{eqC01}
\end{equation}
where $\chi$ is the detector efficiency, $\lambda$ is the light wavelength. The shot noise associated with this current is given by
\begin{equation}
i_{\mathrm{SNL}}=\sqrt{\frac{2ei_{\mathrm{det}}}{\tau}}\ ,\label{eqC02}
\end{equation}
where $\tau$ is the integration time, leading to the following noise equivalent power
\begin{equation}
P_{\mathrm{SNL}}=\frac{i_{\mathrm{SNL}}}{\chi e \lambda/hc}=\sqrt{\frac{2hc}{\chi\lambda\tau}P_{\mathrm{det}}}\ .\label{eqC03}
\end{equation}
The noise on the servo-loop error signal is thus
\begin{equation}
\varepsilon_{\mathrm{SNL}}=\frac{GP_{\mathrm{SNL}}}{\sqrt{2}}=G\sqrt{\frac{hc}{\chi\lambda\tau}P_{\mathrm{det}}}\ ,\label{eqC04}
\end{equation}
where the $1/\sqrt{2}$ factor comes from the fact that only one quadrature is kept after demodulation. Using Eq.\;(\ref{eqB12}), this transforms into the following limit for the detection of the deviation of the optical frequency from resonance:
\begin{equation}
\delta\omega_{\mathrm{SNL}}=\sqrt{\frac{hc}{\lambda\chi\tau}}\sqrt{P_{\mathrm{det}}}\frac{\pi\Delta\nu_{\mathrm{FWHM}}}{2P_0|J_0(\beta)J_1(\beta)|}\ .\label{eqC04N1}
\end{equation}
Using the Sagnac effect scale factor that relates a rotation rate $\dot{\theta}$ into a frequency difference $\Delta\omega_{\mathrm{Sagnac}}$ through
\begin{equation}
\Delta\omega_{\mathrm{Sagnac}}=\frac{2\pi D}{n_0\lambda}\dot{\theta}\ ,\label{eqC05}
\end{equation}
Eq.\,(\ref{eqC04}) translates into the following noise limit on the measurement of a variation of the rotation rate:
\begin{equation}
\delta\dot{\theta}_{\mathrm{SNL}}=\frac{n_0\lambda}{2\pi D}\sqrt{2}\;\delta\omega_{\mathrm{SNL}}\ .\label{eqC06}
\end{equation}
where the factor $\sqrt{2}$ comes from the fact that to measure a variation in the rotation rate, one needs to compare two values of the signal for which the shot noises quadratically add. 
Using Eq.\;(\ref{eqC04}) becomes
\begin{equation}
\delta\dot{\theta}_{\mathrm{SNL}}=\frac{n_0\Delta\nu_{\mathrm{FWHM}}}{D}\sqrt{\frac{\lambda h c}{\chi \tau}}\frac{\sqrt{P_{\mathrm{det}}}}{2\sqrt{2} P_0|J_0(\beta)J_1(\beta)|}\ .\label{eqC07}
\end{equation}
Close to resonance, in critical coupling conditions, the power falling on the detector is due mainly to the two first sidebands, which are fully reflected by the cavity, leading to
\begin{equation}
P_{\mathrm{det}}\simeq2P_0J_1(\beta)^2\ .\label{eqC08}
\end{equation}
Moreover, as shown in Ref. \citenum{Black2001}, the error signal can be maximized by choosing $\beta\simeq1.08\;\mathrm{rad}$, for which  $J_0(\beta)^2\simeq J_1(\beta)\simeq1/2$. In these conditions, Eq.\;(\ref{eqC07}) becomes
\begin{equation}
\delta\dot{\theta}_{\mathrm{SNL}}=\frac{n_0\Delta\nu_{\mathrm{FWHM}}}{D}\sqrt{\frac{\lambda h c}{2\chi \tau P_0}}\ .\label{eqC09}
\end{equation}
which is equivalent to Eq.\;(\ref{eq12}) above.

\section{Kerr Bias}
As shown in Ref.\,\citenum{Lefevre1993}, if the two counter propagating intracavity beams exhibit a power difference $\Delta P$, the Kerr effect creates a difference between the refractive indices seen by the two counter propagating waves that reads
\begin{equation}
\Delta n_{\mathrm{Kerr}}=n_2\frac{\Delta P}{\sigma}\ ,\label{eqD01}
\end{equation}
where $n_2$ is the waveguide effective nonlinear index and $\sigma$ the guided mode area. This leads to a difference between the resonance frequencies of the two counterpropagating modes, given by
\begin{equation}
\Delta \omega_{\mathrm{Kerr}}=2\pi\frac{cn_2}{\lambda n_0}\frac{\Delta P}{\sigma}\ .\label{eqD02}
\end{equation}
Using the gyro scale factor given by Eq.\,(\ref{eqC05}), one obtains the expression of the bias angular velocity induced by the Kerr effect:
\begin{equation}
 \dot{\theta}_{\mathrm{Kerr}}=\frac{cn_2\Delta P}{\sigma D}\ ,\label{eqD03}
\end{equation}
which can be related to the difference $\Delta P_0$ between the powers incident on the cavity in the clockwise and counterclockwise directions using Eqs.\,(\ref{eq09}) and (\ref{eq11}), leading to:
\begin{equation}
 \dot{\theta}_{\mathrm{Kerr}}=\frac{c\,n_2\,\mathcal{F}}{\pi\sigma D}\frac{\Delta P_0}{2}\ .\label{eqD04}
\end{equation}
The extra factor of 2 at the denominator of Eq.\;(\ref{eqD04}) is due to the factor $J_0(\beta)^2$, which relates the carrier power to the total incident power $P_0$ in the Pound-Drever-Hall servo-locking configuration [see Eq.\;(\ref{eqB02})] with $\beta = 1.08\;\mathrm{rad}$.

\label{appendD}

\section{Calculation of Brillouin scattering threshold}
\label{Brillouin}
The threshold pump power for Brillouin scattering in a single pass geometry in a waveguide of length $L$ is approximated in Ref. \citenum{Agrawal2013} by the formula:
\begin{equation}
P_{\mathrm{th}}\simeq 21 \frac{\sigma}{g L_{\mathrm{eff}}}\ ,\label{eqE01}
\end{equation}
where $\sigma$  is the mode area, $L_{\mathrm{eff}}=(1-\mathrm{e}^{-\alpha L})/\alpha$ is the waveguide effective length taking propagation losses into account and $g$ is the Brillouin gain ($g=5\times 10^{-11}\;\mathrm{m}/\mathrm{W}$ for silica). With the SiN waveguides considered in Sec. \ref{applications}, we have $\sigma=33\;\mu\mathrm{m}$. Since $\alpha=0.0736\; \mathrm{m}^{-1}$ for 0.32 dB/km losses (see Table \ref{table01}) and since the cavity diameter $D$ is only few cm large, we  have $L_{\mathrm{eff}}\simeq\pi D$. Assuming for instance a 5 cm diameter cavity and that the low confinement of the mode allows us to consider that it propagates in silica only, we end up with a threshold of 100~W, far above the considered powers. For this calculation, we considered a single pass geometry, because we can choose the cavity free spectral range to make the Brillouin scattered frequency non resonant when the laser beam is resonant, as opposed to a doubly resonant geometry \cite{Norcia2004}. 

\acknowledgments 
We wish to thank J\'er\^ome Bourderionnet, Alfredo De Rossi, and Sylvain Combri\'e for sharing their expertise on integrated photonics technologies with us. This work is supported by the Agence Nationale de la Recherche (Project PHOBAG: ANR-13-BS03-0007) and the European Space Agency (ESA). The work of GF, AR, and FB is performed in the framework of the joint research lab between Thales Research \& Technology and Laboratoire Aim\'e Cotton.

\end{document}